\documentstyle[epsf,prl,twocolumn,floats,aps]{revtex}

\begin{document}

\twocolumn[\hsize\textwidth\columnwidth\hsize\csname @twocolumnfalse\endcsname
\draft
\title{Critical temperature of the superfluid transition in bose liquids}
\author{S.M. Apenko$^{\dagger}$}
\address{I.E. Tamm Theory Division, P.N. Lebedev Physical Institute, Moscow, 117924,
Russia}
\date{\today }
\maketitle

\begin{abstract}
A phenomenological criterion for the superfluid transition is proposed,
which is similar to the Lindemann criterion for the crystal melting. Then
we derive a new formula for the critical
temperature, relating $T_{\lambda}$ to the mean
kinetic energy per particle above the transition. The suppression of the
critical temperature in a sufficiently dense liquid is described as a result
of the quantum decoherence phenomenon. The theory can account for the
observed dependence of $T_{\lambda}$ on density in liquid helium and results
in an estimate $T_{\lambda} \sim 1.1$ K for molecular hydrogen.
\end{abstract}


]

In connection with the search for new bose-condense systems, attention was
again attracted recently to the possible superfluidity in molecular hydrogen
H$_{2}$\cite{Ceperley1,Vorob'ev}, discussed earlier in \cite
{Ginzburg,Akulichev,Maris}. The very possibility crucially depends on the
expected value of the critical temperature $T_{\lambda }$, since normally
molecular hydrogen crystallizes at $\sim 14$ K and some measures should be
taken to keep it in liquid phase at lower temperatures, which may appear
harder to achieve if $T_{\lambda }$ is too low. Hence one needs first to
evaluate $T_{\lambda }$ in supercooled hydrogen. In a more general context
we are faced with the problem of determining $T_{\lambda }$ in a given
quantum liquid. Usually, the first step is to use the well known formula for
the Bose-Einstein condensation temperature in the ideal gas
\begin{equation}
T_{\lambda }\simeq 3.31\frac{\hbar }{m}n^{2/3}  \label{id}
\end{equation}
where $n$ is the density and $m$ is the particle mass ($\kappa _{B}=1$).
Though this formula results e.g. in a rather reasonable estimate of 
$T_{\lambda }$ in liquid helium ($\sim 3$ K instead of the correct value 
$2.17 $ K at the saturated vapor pressure) it still seems unsatisfactory in
case of dense liquids because it cannot account for the observable
dependence of $T_{\lambda }$ on density. According to (\ref{id}) the
critical temperature increases as $n^{2/3}$, while, on the contrary, 
$T_{\lambda }$ in liquid $^{4}$He slightly decreases when the system is
compressed. Certainly, one can argue that it is not the bare mass that
enters in (\ref{id}) but rather an effective one, depending on density \cite
{Feynman}, but there seems to be no simple and general expression for this
effective mass. A calculation of $T_{\lambda }$ based on the Landau
quasiparticle spectrum \cite{Ruvalds} can be applied only when this spectrum
is already known. Qualitatively the observed behavior of $T_{\lambda }$ can
be explained also in the lattice model \cite{Matsubara} but in general it is
not so obvious how to relate the properties of a liquid to those of bosons
on a lattice. For this reasons some new estimate of $T_{\lambda }$ is
required, simple enough to serve as a first approximation, but which can
account for nonmonotonic dependence of the critical temperature of a bose
system on its density. In this Letter we propose such a new general formula,
which relates $T_{\lambda }$ to the mean kinetic energy per particle in the
normal phase --- the quantity which behavior is now well understood \cite
{Ceperley2,Dyugaev,Herwig,Ceperley3}.

The most general approach to superfluidity is, perhaps, to start with the
Feynman path integral expression for the partition function $Z$ for a system
of $N$ interacting bose particles \cite{Feynman}
\begin{equation}
Z=\frac{1}{N!}\sum_{P}\int \prod_{i}d{\bf r}_{i}\int \prod_{i}{\cal D}{\bf r}
_{i}(\tau )\exp (-\frac{1}{\hbar }S)\;,  \label{z}
\end{equation}
\[
S=\int_{0}^{\hbar \beta }\lbrack \frac{m{\bf \dot{r}}_{i}^{2}}{2}
+\sum_{i<j}V({\bf r}_{j}-{\bf r}_{j})\rbrack d\tau
\]
where $V({\bf r}_{i}-{\bf r}_{j})$ is the interparticle interaction
potential and $\beta =1/T$. The integration in (\ref{z}) is over all paths
with ${\bf r}_{i}(0)={\bf r}_{i},$ ${\bf r}_{i}(\hbar \beta )=P{\bf r}_{i}$,
where $P$ is some permutation of $N$ particles, and the sum in (\ref{z}) is
over all such permutations.

At high temperatures only the identity permutation is important, since
particles cannot move far away from their initial positions in `time' $\hbar
\beta $. As $\beta $ increases, a given path ${\bf r}(\tau )$ can spread on
a larger distance, until suddenly it appears possible to end the path at the
position of a neighboring particle. Then, as discussed in detail in the
recent review \cite{CeperleyR}, the rings of exchanges of arbitrary length
are formed, which can be shown to lead to the superfluid behavior. Let us
now consider a system of {\em distinguishable} particles, take arbitrary
particle and evaluate its mean square displacement from initial position 
$\langle (\delta {\bf r)}^{2}\rangle $ in imaginary time. To estimate the
critical temperature, when exchanges can no longer be neglected, we propose
a criterion
\begin{equation}
\langle (\delta {\bf r)}^{2}\rangle \equiv {\frac{1}{\hbar \beta }}
\int_{0}^{\hbar \beta }\langle \lbrack {\bf r}{(\tau )}-{\bf r}{(0)}\rbrack
^{2}\rangle d\tau =\xi a^{2},  \label{cr}
\end{equation}
where $\xi $ is some numerical factor, to be determined later. This
condition merely states that the mean displacement of a given particle in
imaginary time is comparable to the interparticle spacing. The condition 
(\ref{cr}) is similar in spirit to the well known Lindemann criterion in case
of crystal melting and is inspired by the visual representation of the
paths, arising from numerical simulations \cite{CeperleyR}.

In the ideal gas the left hand side of (\ref{cr}) is essentially the square
of the de Broglie thermal wavelength $\lambda _{T}^{2}=\hbar ^{2}/mT$, but
interactions will tend to reduce $\langle (\delta {\bf r)}^{2}\rangle $
(this was observed e.g. in \cite{Cleveland}). There are two main mechanisms
of such a reduction. The first one is related to decoherence due to
interaction with environment. Neighboring particles, in a sense, `measure'
the position of the particle we are looking at, thus reducing its quantum
uncertainty in coordinate space. Such a decoherence phenomenon was discussed
a lot for more than two decades with respect to the transition from the
quantum behavior to the classical one (see e.g. \cite{Kiefer}). The second
mechanism is more typical for crystals or glasses, where particles are
almost localized by potential barriers.

The problem of estimating of $\langle (\delta {\bf r)}^{2}\rangle $ in a
system of interacting particles is still a very complicated one, even if
exchanges are neglected. For a liquid not too close to crystallization it
can be significantly simplified, however, by treating the rest of the system
as some simple fluctuating environment. The simplest choice is the
Caldeira-Leggett model \cite{Caldeira}, which describes interaction of a
particle with a thermal bath of harmonic oscillators. In this model the
particle motion in the imaginary time is governed by the effective action
\begin{eqnarray}
S &=&\int_{0}^{\hbar \beta }d\tau \frac{m{\bf \dot{r}}^{2}}{2}-  \label{s} \\
&&-\frac{1}{4}\int_{0}^{\hbar \beta }d\tau \int_{0}^{\hbar \beta }d\sigma
K(\tau -\sigma )({\bf r}(\tau )-{\bf r}(\sigma ))^{2}  \nonumber
\end{eqnarray}
where the kernel $K(\tau )$ is determined by
\begin{eqnarray}
K(\tau ) &=&(m/\hbar \beta )\sum_{n=-\infty }^{+\infty }\zeta _{n}e^{i\omega
_{n}\tau }\;,  \nonumber \\
\zeta _{n} &=&\frac{1}{m}\int_{0}^{\infty }\frac{d\omega }{\pi }\frac
{I(\omega )}{\omega }\frac{2\omega _{n}}{\omega ^{2}+\omega _{n}^{2}}\;,\;
\label{k}
\end{eqnarray}
where $\omega _{n}=2\pi n/\hbar \beta $ and $I(\omega )$ is the spectral
density of bath oscillators (see e.g. \cite{Grabert}). In our case one can
view (\ref{s}) as a trial action and try to evaluate $K(\tau )$
variationally, but it is easier to directly relate the parameters of the
effective action to some observables of the system (see below). For
quadratic action (\ref{s}) in three dimensions the mean displacement is
\begin{equation}
\langle (\delta {\bf r)}^{2}\rangle =\frac{12}{m\beta }\sum_{n=1}^{\infty }
\frac{1}{\omega _{n}^{2}+\zeta _{n}}  \label{r2}
\end{equation}

At small frequencies it seems natural to expect the ohmic behavior of the
kernel $K(\tau )$, corresponding to the linear friction, when damping is
proportional to the velocity of the particle (cf \cite{Caldeira1}). This is
modeled by $I(\omega )=\gamma m\omega $, where $\gamma $ is the damping
parameter, and $\zeta _{n}=\gamma |\omega _{n}|$. The Caldeira-Leggett model
with such a dissipation kernel is a quantum analog of the standard Langevin
equation with the friction $m\gamma $ and with the white noise random force.
Then the sum in (\ref{r2}) is carried out and
\begin{equation}
\langle (\delta {\bf r)}^{2}\rangle =\frac{6}{\pi }\frac{\hbar }{m\gamma }
(C+\psi (1+\frac{\hbar \gamma \beta }{2\pi }))\;,  \label{rr}
\end{equation}
where $C\simeq 0.577\ldots $ is Euler's constant and $\psi (x)$ is the psi
function (the logarithmic derivative of the gamma function).

Now we have only one parameter, $\gamma $, which describes interactions in
the system. The ideal gas limit is recovered at $\gamma \rightarrow 0$, when
$\langle (\delta {\bf r)}^{2}\rangle \rightarrow \hbar ^{2}\beta /2m$. Then
the criterion (\ref{cr}) results in $T_{\lambda }=(1/2\xi )\hbar
^{2}n^{2/3}/m$. Since this is the ideal gas formula for the critical
temperature (\ref{id}) we conclude that $1/2\xi \simeq 3.31$, i.e.

\begin{equation}
\xi \simeq 0.15  \label{xi}
\end{equation}
Though the criterion (\ref{cr}) is useful only if $\xi $ is some universal
constant we cannot exclude some weak dependence of $\xi $ on density. We
expect that $\xi $ may be slightly smaller than (\ref{xi}) in a system with
short range order since it is more likely for a particle in a liquid to find
a neighbor for exchange at a suitable distance than in the ideal gas, where
density fluctuations are more important \cite{Gruter}.

In the opposite case, when $\hbar \gamma \beta /2\pi \gg 1$ the mean
displacement diverges logarithmically i.e. $\langle (\delta {\bf r)}
^{2}\rangle \sim \ln (\hbar \gamma \beta /2\pi )$. Though $\langle (\delta
{\bf r)}^{2}\rangle $ still tends to infinity as $T\rightarrow 0$, it is
much less, than the de Broglie thermal wavelength of a free particle. This
logarithmic behavior is related to the ohmic spectrum at small frequencies.
Then for the critical temperature at $\gamma \rightarrow \infty $ we have a
very simple formula
\begin{equation}
T_{\lambda }=\alpha \frac{\hbar \gamma }{2\pi }\exp (-\xi \frac{\pi }{6}
\frac{\hbar \gamma }{T_{0}})\;,\qquad T_{0}=\frac{\hbar ^{2}}{m}n^{2/3}
\label{t2}
\end{equation}
where $\alpha =\exp (C)$.

We see now, that the temperature of the superfluid transition crucially
depends on the ratio $\hbar \gamma /T_{0}$. At $\hbar \gamma \ll T_{0}$ the
estimate (\ref{id}) is valid and the critical temperature is essentially 
$T_{0}\sim n^{2/3}$, while at $\hbar \gamma \gg T_{0}$ the critical
temperature of the $\lambda $ transition is exponentially small due to the
decoherence phenomenon. Qualitatively this can be understood as follows: in
the process of exchange particles move through the viscous media and loose
coherence, needed for superfluidity to establish. Since $\gamma $ should
increase with increasing density, the formula obtained do describe the
suppression of $T_{\lambda }$ in sufficiently dense systems.

The approximation of a constant friction would be correct if we dealt with a
heavy Brownian particle, which moves slowly than particles in a liquid. In
real liquids the ohmic spectrum $I(\omega )$ $\sim \omega $ is physical only
at frequencies lower than the collision rate and must have a cutoff at some
frequency $\omega _{c}\sim \gamma $, which now should be taken into account.
One can take e.g. the Drude model for the damping \cite{Grabert}
\begin{equation}
\zeta _{n}={\gamma \omega }_{c}|\omega _{n}|/(\omega _{c}+|\omega _{n}|)
\label{d}
\end{equation}
Apart from the dispersion of the friction coefficient (memory effects) this
implies that the random force acting on a particle is correlated for times
less than $1/\omega _{c}$.

Qualitatively, however, the picture outlined above remains unchanged. If we
e.g. put $\omega _{c}=\gamma $ then again the formula (\ref{t2}) is
recovered but now with $\alpha =\exp (C+\sqrt{3}\pi /9)$.

At low temperatures $\gamma $ can be related to the mean kinetic energy per
particle. The kinetic energy in the Caldeira-Leggett model may be written as
\begin{equation}
K(T)=\frac{3}{2}T+\frac{3}{\beta }\sum_{n=1}^{\infty }\frac{\zeta _{n}}
{\omega _{n}^{2}+\zeta _{n}}  \label{K}
\end{equation}
The sum is easily evaluated for a model (\ref{d}) and e.g. at $\omega
_{c}\rightarrow \infty $ (purely ohmic limit) we obtain
\begin{equation}
K(T)=\frac{3}{2}T+\frac{3\hbar \gamma }{2\pi }\left[ \ln \frac{\hbar \omega
_{c}\beta }{2\pi }-\psi (1+\frac{\hbar \gamma \beta }{2\pi })\right]
\label{k1}
\end{equation}
This very expression was derived earlier \cite{Dyugaev} for liquid $^{4}$He
above the $\lambda $ point by taking the velocity autocorrelation function
to be of a simple exponential form with damping $\gamma $ (denoted in \cite
{Dyugaev} by $\omega _{0}$), just as for the Brownian particle. At $T\gg
\hbar \gamma $ the kinetic energy tends to $\frac{3}{2}T$, while at low
temperatures $K(T)$ has a finite limit, denoted hereafter by $\dot{K}$. This
is quite a general behavior, valid in the Debye model as well \cite{Andreani}
. Strictly speaking, $K$ is the kinetic energy just above $T_{\lambda }$,
since exchanges are not taken into account, but e.g. in helium the
difference between $K\sim 16$ K and the real zero-point energy is $\sim 1.5$
K and will be neglected here. For the particular model (\ref{k1}) we have 
$K=(3\hbar \gamma /2\pi )\ln (\omega _{c}/\gamma )$, i.e., as was already
mentioned in \cite{Dyugaev} (see also \cite{Golubev}), the frequency $\gamma
$ up to a logarithmic factor coincides with $K$. Thus we conclude that in
general case at low temperatures
\begin{equation}
\hbar \gamma \sim K  \label{g}
\end{equation}
For different $I(\omega )$ the proportionality coefficient is actually
cutoff dependent and e.g. for the model (\ref{d}) with $\omega _{c}=\gamma $
one has $\hbar \gamma =\sqrt{3}K$. Since in the formula (\ref{t2}) the
factor $\alpha $ also depends on a high frequency behavior of $I(\omega )$
we can finally write
\begin{equation}
T_{\lambda }=A\,K\exp (-B\,\frac{K}{T_{0}})  \label{tf}
\end{equation}
for $K\gg T_{0}$, where $A$ and $B$ are some model dependent constants. For
the Caldeira-Leggett model with damping kernel (\ref{d}) at $\omega
_{c}=\gamma $ we have
\begin{eqnarray}
A &=&(\sqrt{3}/2\pi )\exp (C+\sqrt{3}\pi /9)\simeq 0.899,  \nonumber \\
B &=&(\sqrt{3}\pi /6)\,\xi \simeq 0.907\,\xi  \label{a}
\end{eqnarray}
The physical meaning of (\ref{tf}) is clear: if $K$, which at $T\ll K$ may
be viewed as an effective `internal' temperature of the system \cite{Dyugaev}, 
is much larger than the transition temperature in the ideal gas, the
superfluidity is suppressed.

Let us now compare the formula (\ref{tf}) with experimental data for liquid
helium. We need then some explicit expression for $K(n)$. There are
different estimates of zero-point energy in helium (see e.g. \cite
{London,Dyugaev}). All of them are in general consistent both with the
experimental data \cite{Herwig} and with results of Path Integral Monte
Carlo (PIMC) calculations \cite{Ceperley2,Ceperley3}. The kinetic energy
increases with density due to the repulsion core in interatomic potential.
Here we shall use the London's formula
\begin{equation}
K=\frac{2\pi \hbar ^{2}d}{m(a-0.891d)^{2}(a+0.713d)}  \label{L}
\end{equation}
where $a=n^{-1/3}$, $d\simeq 2.4$ \AA\ \cite{London}. This formula is simple
and transparent being an interpolation between low density limit $K\sim
\hbar ^{2}dn/m$ of the energy of a gas of hard spheres of radius $d$, and a
quantum mechanical estimate $K\sim \hbar ^{2}/m(a-d_{0})^{2}$ with 
$d_{0}\sim d$ in the high density limit.

\begin{figure}[htp]
\epsfxsize=3.375in
\centerline{\epsffile{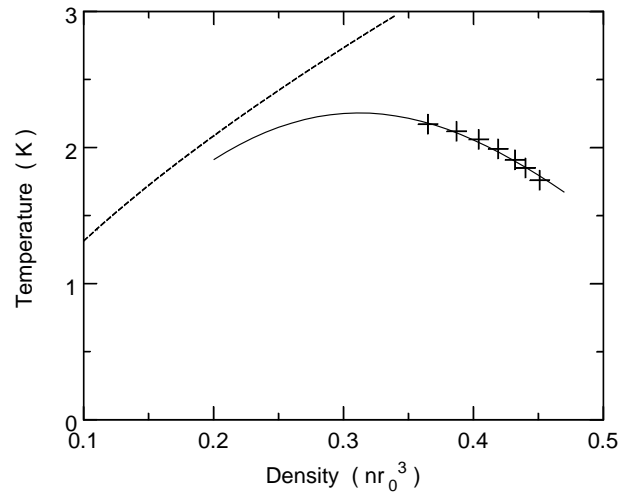}}

\caption{
Temperature of the superfluid transition vs the reduced density $n^{\ast}=nr_{0}^{3}$ ($r_{0}=2.556$ \AA). Solid line is the theory at
$\xi=0.12$, crosses
denotes experimental data for helium, dashed line corresponds
to the ideal gas.
}

\label{fig}
\end{figure}

With this expression for $K$ the formula (\ref{tf}) with coefficients (\ref
{a}) at $\xi =0.15$ results in $T_{\lambda }\simeq 1.35$ K, which is lower
than the experimental value of $2.17$ K. We may recall, however, that $\xi $
may be smaller in dense system than in the ideal gas, and try to fit (\ref
{tf}) to experimental data treating $\xi $ as an adjustible parameter. The
result of such a fit is shown in Fig. 1. Here the critical temperature is
shown as a function of the reduced density $n^{\ast }=nr_{0}^{3}$, where 
$r_{0}=2.556$ \AA\ is the length parameter of the Lennard-Jones interatomic
potential for helium. Experimental values of $T_{\lambda }$ for liquid
helium, taken from the very accurate empirical expression for the $\lambda$-line 
\cite{Kierstead}, are shown by crosses. The dashed curve represents
the ideal gas formula (\ref{id} ), while the solid one is the best fit of 
$T_{\lambda }$ from (\ref{tf}) and (\ref{a}) to experiment, which corresponds
to $\xi \simeq 0.12$. This value is only slightly smaller than the ideal gas
limit $\xi =0.15$. Given the simplicity of the assumptions, the agreement of
the theory with experiment is quite satisfactory. Since $BK/T_{0}\sim 1$ at 
$n^{\ast }\sim 0.2$, we can not expect equation (\ref{tf}) to be
quantitatively valid at such a low density, but still it is clear that the
theory really can describe the crossover from the ideal gas behavior to the
observed dependence of $T_{\lambda }$ on the density.

Now, as far as the molecular hydrogen is concerned, we may take the London's
formula (\ref{L}) with $d\simeq 2.7$ \AA\ \cite{Wilks}, which corresponds to
a stronger interparticle potential. Assuming the numerical coefficients $A$
and $B$ to be the same as in helium (with $\xi \simeq 0.12$) we obtain from 
(\ref{tf}) $T_{\lambda }\simeq 1.1$ K for density $n\simeq 26$ nm$^{-3}$.
Even if we reduced the density to that of helium i.e. $n\simeq 22$ nm$^{-3}$
the critical temperature would be only $\sim 2.1$ K. This is much less
optimistic than the original estimate $6\div 8$ K \cite{Ginzburg}, based on
the ideal gas formula (\ref{id}), though consistent with later estimates
\cite{Maris} and with the PIMC analysis of finite hydrogen clusters
\cite{Ceperley4}. The value $\sim 1$ K, obtained recently for hydrogen films
with impurities \cite{Ceperley1} is also of the same order.

In conclusion, we have obtained a general estimate for the critical
temperature of the superfluid transition in a bose liquid. Starting from the
phenomenological Lindemann-like criterion (\ref{cr}) for the transition and
modelling the decoherence effect, which suppresses $T_{\lambda }$ in liquid,
by the Caldeira-Leggett model, we arrive at a simple expression (\ref{tf}),
relating $T_{\lambda }$ to the kinetic energy $K$ (which is essentially the
zero-point energy per particle). The fast increase of $K$ with density due
to the repulsion core in interatomic potential accounts then for the
reduction of $T_{\lambda }$ in dense systems, when $K\gg T_{0}\sim \hbar
^{2}n^{2/3}/m$, as observed e.g. in liquid helium. Using the  London's 
interpolation formula for $K(n)$ and one fitting parameter the model
can even quantitatively describe experimental behavior of $T_{\lambda }$. 

Our approach suggests that, in a liquid, contrary to weakly nonideal gas, $T_{0}$ 
may not correspond to any characteristic temperature. The temperature
where quantum effects become important is of the order of $K$, and is much
larger than $T_{0}$ \cite{Dyugaev}, while $T_{\lambda }$ at high densities
is smaller than $T_{0}$, and the difference increases with density. The
error of using the ideal gas formula (\ref{id}) is not so much for helium,
but may be of importance in molecular hydrogen (which is more dense, due to
stronger interaction), where our formula gives much smaller value of $T_{\lambda }$.

An important question, which remains open within the phenomenological
approach, concerns the universality of the numerical constants in (\ref{tf}
). Though the parameter $\xi $ (analogous to the Lindemann ratio in case of
crystal melting) is to some extent fixed by the ideal gas limit (\ref{xi})
further work is needed to clarify its possible dependence on system
parameters.

I am very grateful to V.L.Ginzburg for attracting my attention to the
problem, and to S.P. Malyshenko and V.S. Vorob'ev, D.S. Golubev, P.I.
Arseyev and V.V. Losyakov for fruitful discussions.

\end{document}